\documentclass[prd,floatfix,onecolumn,amsmath]{revtex4-1}
\usepackage{graphicx,color,dcolumn,booktabs,bm}
\usepackage{subfigure}
\usepackage{longtable}
\usepackage{amssymb}
\usepackage{indentfirst}
\usepackage{epsfig}
\usepackage{feynmf}   
\usepackage{epstopdf}   
\usepackage{slashed}  
\usepackage{cases}
\usepackage[pdfpages]{xcolor}
\usepackage{pgf}
\definecolor{maroon}{RGB}{139,25,150}
\usepackage{multirow}
\usepackage{physics}
\usepackage{float}
\usepackage[colorlinks, citecolor=magenta,anchorcolor=red,menucolor=blue, linkcolor=blue,filecolor=red,runcolor=blue,urlcolor=magenta,frenchlinks=blue]{hyperref}

\begin{document}

	\preprint{}
	
\title{\color{maroon}{Can we determine the exact size of the nucleon?: \\
A comprehensive study of different radii}}

\author{The MMGPDs Collaboration:\\
Muhammad Goharipour$^{1,2}$}
\email{muhammad.goharipour@ipm.ir}
\thanks{Corresponding author}

\author{Fatemeh Irani$^{3}$}
\email{f.irani@ut.ac.ir}

\author{M.~H.~Amiri$^{3}$}
\email{mh.amiri79@ut.ac.ir}

\author{H.~Fatehi$^{3}$}
\email{apranik.fatehi@cern.ch}

\author{Behnam Falahi$^{3}$}
\email{behnam.falahi@ut.ac.ir}

\author{A.~Moradi$^{3}$}
\email{a.moradi1992@ut.ac.ir}

\author{K.~Azizi$^{3,4,2}$}
\email{kazem.azizi@ut.ac.ir}

\affiliation{
$^{1}$School of Physics, Institute for Research in Fundamental Sciences (IPM), P.O. Box  19395-5531, Tehran, Iran\\
$^{2}$School of Particles and Accelerators, Institute for Research in Fundamental Sciences (IPM), P.O. Box 19395-5746, Tehran, Iran\\
$^{3}$Department of Physics, University of Tehran, North Karegar Avenue, Tehran 14395-547, Iran\\
$^{4}$Department of Physics, Dogus University, Dudullu-\"{U}mraniye, 34775 Istanbul, T\"urkiye }

\date{\today}

\begin{abstract}

The concept of nucleon radii plays a central role in our understanding of the internal structure of protons and neutrons, providing critical insights into the non-perturbative regime of quantum chromodynamics (QCD). While the charge radius is often interpreted as the ``size" of the nucleon, this interpretation is an oversimplification that overlooks the multifaceted nature of nucleon structure. This paper provides a comprehensive overview of the different nucleon radii, including the charge and magnetic radii, the axial radius, and the emerging concepts of mechanical and mass radii. We discuss the definitions as well as the experimental, theoretical and phenomenological determinations of these radii, highlighting their distinct physical origins and implications. By synthesizing recent experimental results and theoretical advancements, we emphasize that each radius reflects a specific aspect of the nucleon's internal structure, such as its electric charge distribution, magnetic properties, weak interactions, or internal mechanical stress. In particular, we address the common but misleading interpretation of the proton radius as a simple measure of its size, underscoring the nuanced and context-dependent nature of nucleon radii. Through this exploration, we aim to clarify the roles of these radii in characterizing nucleon structure and to identify open questions that remain to be addressed. This work contributes to a deeper understanding of the nucleon and its significance in the broader context of particle and nuclear physics.
	
\end{abstract}

\maketitle

\section{Introduction}\label{sec:one} 

The nucleon, comprising protons and neutrons, is a fundamental building block of atomic nuclei and plays a central role in our understanding of nuclear and particle physics. Despite decades of research, the internal structure of nucleons remains a subject of intense investigation, as it provides critical insights into the non-perturbative regime of quantum chromodynamics (QCD). One of the key observables that characterize the spatial distribution of nucleons' internal properties is their radii. These radii are not only essential for understanding the nucleon's structure but also have profound implications for precision tests of the Standard Model. In particular, precise measurements of these radii serve as stringent tests for theories such as chiral effective field theory, light-cone QCD sum rules (LCSR), and lattice QCD.

Nucleon radii are defined in terms of various physical observables, each reflecting a different aspect of the nucleon's internal structure. Among them, the {\it charge} and {\it magnetic} radii are most common~\cite{Karr:2020wgh,Pohl:2010zza,Gao:2021sml,Xiong:2023zih,Meissner:2022rsm,Cui:2022fyr,Ridwan:2023ome}.
The charge radius, extracted from electron scattering and spectroscopy experiments, describes the spatial distribution of the electric charge within the nucleon. The magnetic radius, on the other hand, characterizes the distribution of the nucleon's magnetic moment. These electromagnetic radii have been extensively studied, yet discrepancies, such as the proton charge radius puzzle, highlight the complexity of interpreting these quantities~\cite{Lorenz:2012tm,Pohl:2013yb,Bernauer:2014cwa,Carlson:2015jba,Hill:2017wzi,Peset:2021iul,Lin:2023fhr,Lumpay:2025btu}.

Beyond electromagnetic properties, the {\it axial} radius is associated with the distribution of the nucleon's weak charge, as probed in neutrino scattering and beta decay experiments. This radius is crucial for understanding the nucleon's role in weak interactions and has implications for precision measurements in neutrino physics~\cite{Bernard:2001rs,MiniBooNE:2010bsu,MINERvA:2023avz,MiniBooNE:2010xqw,Butkevich:2013vva,Meyer:2016oeg,Hill:2017wgb}.

In addition to these traditional radii, recent advances in theory and experiment have introduced the concept of {\it mechanical} and {\it mass} radii~\cite{Polyakov:2018zvc,Lorce:2018egm,Burkert:2023wzr}. The mechanical radius describes the spatial distribution of pressure and shear forces within the nucleon, providing a unique perspective on the dynamics of the strong force. The mass radius, derived from the energy-momentum tensor (EMT), reflects the distribution of the nucleon's mass, which is deeply connected to the origin of mass in QCD. These radii offer complementary insights into the nucleon's structure and are essential for a comprehensive understanding of its properties.

Considering the wide range of values obtained for these radii from various experimental, theoretical, and phenomenological studies, the question arises: {\it Can we interpret these radii as the size of the nucleon?} It appears that interpreting nucleon radii as the ``size" of the nucleon is a nuanced and context-dependent topic. While these radii provide measures of spatial distributions associated with different properties of the nucleon-such as charge, magnetism, weak interactions, mechanical stress, and mass distribution-they do not all correspond to a single, unified concept of size. To be more precise, each radius offers a measure of the nucleon's spatial extent in the context of a specific physical property, but there is no single, universally agreed-upon definition of the nucleon's size. For example, the charge radius is often interpreted as the size of the nucleon in terms of its electric charge distribution~\cite{Pohl:2010zza,Karr:2020wgh}, but this interpretation is open to criticism given the multifaceted nature of nucleon structure and the differences between various radii~\cite{Liu:1998um,Kaiser:2024vbc,Petrov:2024hwj}. Thus, while each radius provides valuable insights into the nucleon's internal structure, they collectively highlight the complexity of defining a single ``size" for the nucleon. 

This paper aims to provide a detailed overview of the different nucleon radii, including their definitions, experimental determinations, and theoretical interpretations. We will discuss the charge and magnetic radii, the axial radius, and the emerging concepts of mechanical and mass radii. By synthesizing recent experimental results and theoretical developments, we seek to highlight the progress made in understanding nucleon structure and to identify open questions that remain to be addressed. In particular, we aim to address the common but oversimplified interpretation of the proton radius as the ``size" of the proton, emphasizing that this concept is more nuanced and context-dependent than often assumed. Through this exploration, we hope to contribute to a deeper understanding of the nucleon and its role in the broader context of particle and nuclear physics.

\section{The charge and magnetic radii}\label{sec:two}

The charge and magnetic radii are among the most fundamental observables used to characterize the internal structure of nucleons. The charge radius describes the spatial distribution of the electric charge within the nucleon, while the magnetic radius reflects the distribution of its magnetic moment, which arises from the motion of quarks and gluons. These radii provide critical insights into the electromagnetic structure of protons and neutrons and serve as key benchmarks for testing theoretical models of QCD. Historically, the proton's charge radius has been a subject of intense scrutiny~\cite{Karr:2020wgh,Gao:2021sml,Xiong:2023zih,Pohl:2010zza,Meissner:2022rsm,Cui:2022fyr,Ridwan:2023ome}, particularly in light of the ``proton radius puzzle," which revealed discrepancies between measurements obtained from electron scattering and laser spectroscopy of hydrogen atoms~\cite{Lorenz:2012tm,Pohl:2013yb,Bernauer:2014cwa,Carlson:2015jba,Hill:2017wzi,Peset:2021iul,Lin:2023fhr,Lumpay:2025btu,Jentschura:2010ha,Carlson:2012pc,Barger:2010aj,Karr:2012mfa,Wang:2013fma,Onofrio:2013fea,Miller:2018ybm,Dahia:2023urs}. Similarly, the magnetic radius, though less commonly discussed, offers valuable information about the nucleon's internal dynamics and its response to external magnetic fields. In this section, we review the results of the experimental techniques used to determine the charge and magnetic radii, including electron scattering and spectroscopy. We also discuss the results of different theoretical models and phenomenological frameworks  employed to determine and interpret these radii. 

The charge and magnetic radii of nucleons are determined through a variety of experimental techniques, each with its own strengths, limitations, and sources of systematic uncertainty. One of the most direct methods is elastic electron scattering, where high-energy electrons are scattered off nucleons, and the differential cross-section is measured~\cite{Gao:2021sml,Xiong:2023zih}. By analyzing the scattering data, the electric and magnetic form factors (FFs), $ G_E(t) $ and $ G_M(t) $, also called the ``Sachs" FFs, or their ratios can be extracted as functions of the momentum transfer squared, $ t=-Q^2 $~\cite{Gao:2003ag,Qattan:2004ht,Perdrisat:2006hj,Crawford:2006rz,A1:2010nsl,A1:2013fsc,Ye:2017gyb,Xiong:2019umf,Christy:2021snt}. The mean squared of the charge and magnetic radii are then obtained from the slopes of these form factors at $ Q^2\rightarrow 0 $,
\begin{align}
\left<r_{jE}^2\right>= \left.  6 \dv{G_E^j}{t} \right|_{t=0} \,, ~~~~~~
\left<r_{jM}^2\right>= \left.  \frac{6}{\mu_j} \dv{G_M^j}{t} \right|_{t=0}\,,
\label{Eq1}
\end{align}
where  $ j=p,n $ denotes the proton and neutron, respectively, and $ \mu_j $ is the magnetic moment of the nucleon. While electron scattering provides precise measurements of the proton's charge and magnetic radii, its application to the neutron is more challenging due to the neutron's lack of a net electric charge. For the neutron, the measurements of the double polarization observables in quasi-elastic electron scattering from polarized deuterium or $ ^3 $He targets
is often used to determine the neutron electric FF, $ G_E^n $, though this method introduces additional theoretical uncertainties related to nuclear corrections and was not able to access $ G_E^n $ at a sufficiently low $ Q^2 $ range~\cite{Atac:2021wqj}. It should be also noted that radii are not fundamental quantities in quantum field theory (QFT). The conventional definition of $ \left<r^2\right> $ in Eq.~(\ref{Eq1}), rooted in non-relativistic form factor expansions at low $ Q^2 $, is inadequate for relativistic systems~\cite{Miller:2018ybm,Miller:2007uy,Chen:2022smg,Chen:2023dxp}. A more rigorous framework is provided by impact-parameter-dependent GPDs~\cite{Ji:1998pc,Burkardt:2000za,Burkardt:2002hr,Singireddy:2025biy}, which avoid the ambiguities of a universal radius definition. These QFT-based approaches highlight the limitations of interpreting  $ \left<r^2\right> $ as a strict measure of spatial extent.

Since the charge radius of the proton ($ r_{pE} = 0.74 \pm 0.24 $ femtometers (fm)) was first determined by Robert Hofstadter and collaborators in the 1950s at Stanford University using electron-proton ($ ep $) elastic scattering experiments~\cite{Hofstadter:1955ae,Mcallister:1956ng}, numerous experimental efforts have been undertaken to refine the electromagnetic radii of the nucleon through scattering techniques. Let's start with the the proton charge and magnetic radii, $ r_{pE} $ and $ r_{pM} $, as extracted from scattering experiments since 2010. Bernauer \textit{et al.} determined these radii using an unpolarized $ ep $ elastic scattering experiment at the Mainz accelerator facility MAMI~\cite{A1:2010nsl,A1:2013fsc}. In contrast, Zhan \textit{et al.}~\cite{Zhan:2011ji} performed a high-precision measurement of the polarization transfer in $ ep $ elastic scattering, employing a recoil proton polarimeter. They extracted the proton charge and magnetic radii by updating the global fit of the proton form factors (FFs) performed by Arrington \textit{et al.}~\cite{Arrington:2007ux}, incorporating these measurements along with additional data on the proton FFs ratio from Jefferson Lab. Using the technique of initial-state radiation (ISR), Mihovilovic \textit{et al.}~\cite{Mihovilovic:2016rkr,Mihovilovic:2019jiz} measured unpolarized $ ep $ elastic scattering at lower values of $ Q^2 $, enabling an extraction of the proton charge radius. Xiong \textit{et al.}~\cite{Xiong:2019umf} employed a magnetic-spectrometer-free method combined with a windowless hydrogen gas target in the PRad experiment at Jefferson Lab to measure the electric form factor $ G_E $. This innovative approach overcame several limitations of previous $ ep $ experiments, allowing for measurements at very small $ Q^2 $ values and providing a more precise determination of the proton charge radius. In Ref.~\cite{Epstein:2014zua}, Epstein \textit{et al.} extracted both the proton and neutron magnetic radius in a model independent way from the electron-proton scattering data in addition to the electron-neutron scattering and $ \pi\pi \rightarrow N\bar{N} $ data. Chwastowski \textit{et al.} indicated that the two-photon exclusive production of lepton pairs at the Electron–Ion Collider can provide unique measurements of the proton elastic electromagnetic FFs and hence the nucleon's radii~\cite{Chwastowski:2022fzk}.

In the case of the neutron, the determination of the mean-squared charge and magnetic radii is more challenging due to the absence of a free neutron target, which introduces significant experimental and theoretical uncertainties in electron scattering methods. Despite these challenges, several techniques have been developed to extract the neutron's electromagnetic radii. In addition to the Epstein \textit{et al.} determination described above, for instance, low-energy neutron scattering experiments have been conducted using electrons bound in diamagnetic atoms~\cite{Krohn:1973re,Koester:1995nx,Kopecky:1997rw}. Krohn \textit{et al.}~\cite{Krohn:1973re} utilized Ne, Ar, Kr, and Xe atoms, while Koester \textit{et al.}~\cite{Koester:1995nx} and Kopecky \textit{et al.}~\cite{Kopecky:1997rw} focused on heavier atoms such as Bi and Pb. An alternative approach, demonstrated by Atac \textit{et al.}~\cite{Atac:2021wqj}, involves extracting the neutron electric form factor $ G_E^n $ from high-precision measurements of the quadrupole and magnetic dipole nucleon-to-delta ($ N\rightarrow \Delta $) transition form factors (TFFs). This method provides access to $ G_E^n $ at lower $ Q^2 $ values, offering a complementary pathway to determine the neutron's charge radius. Additionally, Heacock \textit{et al.}~\cite{Heacock:2021btd} employed neutron pendell\"{o}sung measurements, a sophisticated technique in neutron diffraction studies, to investigate the internal structure of perfect crystals and determine the neutron's mean square charge radius. These diverse approaches highlight the ingenuity required to overcome the inherent difficulties in studying the neutron's electromagnetic properties. It is worth noting in this context that the parity-violating electron scattering (PVES) can also be used as a complementary tool for precise determination of neutron's charge distribution in nuclei~\cite{Ban:2010wx,Abrahamyan:2012gp,Horowitz:2013wha,Souder:2015mlu}.

Another powerful experimental technique to determine the proton charge radius, $ r_{pE} $ , is laser spectroscopy of hydrogen atoms, which measures the energy levels of atomic electrons (or muons) bound to a proton~\cite{Pohl:2010zza,Karr:2020wgh}. The Lamb shift, a small difference in energy levels due to quantum electrodynamics (QED) effects, is sensitive to the proton's charge radius. Therefore, high-precision spectroscopic measurements, combined with accurate QED calculations, can determine $ r_{pE} $ with remarkable precision. However, this method led to the proton radius puzzle when measurements using muonic hydrogen (where an electron is replaced by a heavier muon) yielded a significantly smaller $ r_{pE} $ compared to traditional electron-based spectroscopy~\cite{Lorenz:2012tm,Pohl:2013yb,Bernauer:2014cwa,Carlson:2015jba,Hill:2017wzi,Peset:2021iul,Lin:2023fhr,Lumpay:2025btu}. This discrepancy has spurred extensive theoretical and experimental efforts to resolve the puzzle, including improved QED calculations and new experimental techniques~\cite{Gao:2021sml,Lumpay:2025btu}. For ordinary hydrogen spectroscopic measurements of the proton charge radius, one can refer to the works of Beyer \textit{et al.}~\cite{Beyer:2017gug}, Fleurbaey \textit{et al.}~\cite{Fleurbaey:2018fih}, Bezginov \textit{et al.}~\cite{Bezginov:2019mdi}, Grinin \textit{et al.}~\cite{Grinin:2020txk}, and Brandt \textit{et al.}~\cite{Brandt:2021yor}, all of which were performed after the release of the first muonic hydrogen spectroscopic determination of the proton charge radius by Pohl \textit{et al.}~\cite{Pohl:2010zza}. This precise determination of $ r_{pE} $ from muonic hydrogen spectroscopy was later updated by Antognini \textit{et al.}~\cite{Antognini:2013txn}, resulting in a more precise and slightly smaller value.

Besides the direct experimental extractions of the nucleon charge and magnetic radii, there have been various analyses of the elastic scattering data~\cite{Ridwan:2023ome,Lorenz:2012tm,Kelly:2002if,Lorenz:2014vha,Kraus:2014qua,Higinbotham:2018jfh,Zahra:2022tkq,Pacetti:2021fji,Cui:2021skn,Sick:2017aor,Horbatsch:2015qda,Lin:2021umz,Horbatsch:2019wdn,Hagelstein:2018zrz,Borisyuk:2020pxo,Borah:2020gte,Filin:2019eoe,Filin:2020tcs,Barcus:2019skg,Vaziri:2023xee,Griffioen:2015hta,Alarcon:2018zbz,Hill:2010yb,Zhou:2018bon,Cui:2021vgm,Sick:2018fzn,Higinbotham:2015rja,Belushkin:2006qa,Borisyuk:2009mg,Lorenz:2014yda,Graczyk:2014lba,Alarcon:2020kcz,Lee:2015jqa,Arrington:2015ria,Horbatsch:2016ilr,Lin:2021umk,Lin:2021xrc,Goharipour:2024mbk,Atac:2020hdq,Gramolin:2021gln,Hayward:2018qij,Paz:2020prs,Atoui:2023nrn,Albloushi:2024wxc} aimed at resolving the discrepancies observed between different experimental techniques. These studies employ diverse methodologies, including conformal mapping, dispersive approaches, Bayesian frameworks, and statistical sampling techniques, to extract the nucleon radii, particularly the proton charge radius, from experimental data~\cite{Gao:2021sml}. Most analyses rely on high-precision data from the Mainz A1 experiment~\cite{A1:2010nsl,A1:2013fsc} and, more recently, the PRad experiment~\cite{Xiong:2019umf}. These datasets provide a robust foundation for extracting the proton charge radius with reduced uncertainties. Many analyses, such as those by Lorenz \textit{et al.}~\cite{Lorenz:2012tm,Lorenz:2014vha,Lorenz:2014yda}, Griffioen \textit{et al.}~\cite{Griffioen:2015hta}, Alarcón \textit{et al.}~\cite{Alarcon:2018zbz,Alarcon:2020kcz}, and Lin \textit{et al.}~\cite{Lin:2021umk,Lin:2021xrc}, yield $ r_{pE} $ values consistent with the smaller muonic hydrogen results ($ \sim 0.84 $ fm). These studies often incorporate advanced theoretical constraints, such as dispersive improvements, chiral perturbation theory ($ \chi $PT), and Two-Photon-Exchange (TPE) corrections, to ensure precision. On the other hand, many analyses, such as those by Hill and Paz~\cite{Hill:2010yb}, Graczyk and Juszczak~\cite{Graczyk:2014lba}, Zhou \textit{et al.}~\cite{Zhou:2018bon}, Cui \textit{et al.}~\cite{Cui:2021vgm,Cui:2021skn}, and Paz~\cite{Paz:2020prs}, employ model-independent techniques (e.g., conformal mapping, Bayesian methods, or the Schlessinger Point Method) to minimize biases from assumed functional forms. These analyses also often lead to $ r_{pE} $ values consistent with the smaller muonic hydrogen results. However, some studies, such as Lee \textit{et al.}~\cite{Lee:2015jqa}, Sick~\cite{Sick:2018fzn}, Arrington and Sick~\cite{Arrington:2015ria}, and Gramolin and Russell~\cite{Gramolin:2021gln}, report larger values, closer to older electron scattering results.

The differences between these analyses often stem from variations in methodology, data selection, and treatment of systematic uncertainties. For example, some studies, like Griffioen \textit{et al.}~\cite{Griffioen:2015hta}, inflate systematic uncertainties to account for normalization issues, while others, like Alarcón \textit{et al.}~\cite{Alarcon:2018zbz,Alarcon:2020kcz}, incorporate theoretical uncertainties from spectral functions or $ \chi $PT. As another example, Sick and Trautmann~\cite{Sick:2017aor} argued that neglecting higher moments of the charge density distribution in some analyses leads to smaller radius values, and Horbatsch \textit{et al.}~\cite{Horbatsch:2016ilr} addressed this by fixing higher moments using $ \chi $PT. It is also worth noting in this context the following analyses. Horbatsch~\cite{Horbatsch:2019wdn} analyzed the PRad data by taking the logarithm to yield a $ Q^2 $-dependent radius function. Atac \textit{et al.}~\cite{Atac:2020hdq} extracted both the proton and neutron charge radii from a global analysis of the proton and neutron elastic form factors by carrying out a flavor decomposition of these FFs under charge symmetry in the light-cone frame formulation. Gramolin and Russell~\cite{Gramolin:2021gln} proposed a novel method for extracting $ r_{pE} $, which relates the radius of the proton to its transverse charge density, the two-dimensional Fourier transform of the Dirac form factor. Filin \textit{et al.}~\cite{Filin:2019eoe,Filin:2020tcs} extracted the neutron charge radius through a comprehensive analysis of the deuteron charge and quadrupole form factor data in chiral effective field theory. Atoui \textit{et al.}~\cite{Atoui:2023nrn} determined $ r_{pE} $ through a global analysis of proton electric form factor data based on an integral method.

The MMGPDs Collaboration~\cite{Goharipour:2024mbk} determined the charge and magnetic radii of both the proton and neutron through a simulation analysis of a wide range of elastic scattering data, considering the relation between the Pauli and Dirac  FFs (and hence the Sachs FFs) and the generalized parton distributions (GPDs) at zero skewness. While GPDs offer a powerful framework for investigating nucleon structure through their unified description of spatial and momentum-space distributions, their application to radius extraction may need further consideration. Actually, the integration over the momentum fraction $x$ may introduce additional model dependence compared to direct form factor analyses, potentially affecting the precision of radius determinations. Nevertheless, note that the extracted GPDs leads to consistent results for other observables and quantities such as the electromagnetic, axial, gravitational, and transition FFs~\cite{Hashamipour:2022noy,Irani:2023lol,Goharipour:2024atx}. Finally, note that the extracted radius values can be sensitive to the $ Q^2 $ range of the data included, as mentioned before. For example, Ridwan and Mart~\cite{Ridwan:2023ome}, Higinbotham \textit{et al.}~\cite{Higinbotham:2015rja}, Zhou \textit{et al.}~\cite{Zhou:2018bon}, and Pacetti and Tomasi-Gustafsson~\cite{Pacetti:2021fji} highlight the impact of data selection on the final results. The values extracted by Zhou \textit{et al.}~\cite{Zhou:2018bon} for the proton charge radius range from 0.83 to 0.88 fm, while the mixed result is $ r_{pE} = 0.8550^{+0.0037}_{-0.0036} $ fm. As another example, Hill and Paz~\cite{Hill:2010yb} indicated that adding the neutron and $ \pi\pi $ data in the analysis leads to a decrease in the uncertainty of the proton charge radius. In addition, Barcus \textit{et al.}~\cite{Barcus:2019skg} indicated how analytic choices, such as the function that is being used or even the binning of the data, can affect the extraction of electromagnetic FFs from elastic electron scattering cross section data and hence the extracted proton charge radius.

In addition to experimental measurements and phenomenological analyses, a variety of  theoretical approaches have been developed to calculate the electromagnetic FFs or the charge and magnetic radii of the nucleon~\cite{Yennie:1957skg,Ramalho:2011pp,Trinhammer:2021irt,CSSM:2014knt,Shanahan:2014cga,QCDSF:2017ssq,Alexandrou:2017ypw,Alexandrou:2020aja,Alexandrou:2018sjm,Hasan:2017wwt,Park:2021ypf,Ishikawa:2021eut,Shintani:2018ozy,Tsuji:2023llh,Jang:2019jkn,Djukanovic:2021cgp,Djukanovic:2023jag,Djukanovic:2023beb,Syritsyn:2023pmn,Abidin:2009hr,Brodsky:2014yha,Sufian:2016hwn,Gutsche:2019jzh,Contreras:2021epz,Mamo:2021jhj,Ahmady:2021qed,Xu:2021wwj,Mondal:2019jdg,Maris:2003vk,Eichmann:2009qa,Eichmann:2016yit,Burkert:2017djo,Ding:2022ows,Ferreira:2023fva,Barabanov:2020jvn,Yao:2024uej,Gasser:1987rb,Kubis:2000aa,Alarcon:2012nr,Bauer:2012pv,Liu:2013fda,Flores-Mendieta:2015wir,He:2017viu,HillerBlin:2017syu,Yang:2020rpi,Bar:2021crj,Alvarado:2023loi,Aung:2024qmf,Berger:2004yi,Julia-Diaz:2003sdv,Sharma:2013mfa,Plessas:2015mpa,Balitsky:1985ag,Lenz:2003tq,Castillo:2003pt,Braun:2006hz,Aliev:2008cs,Anikin:2013aka,Brodsky:2022kef,Gao:2022osh,Xing:2023uhj,Ramalho:2023hqd,Faessler:2009tn,Kuzmin:2024ozz}. These approaches are rooted in QCD, but they employ different strategies to address the non-perturbative nature of QCD at low energies. One of the most prominent methods is lattice QCD, which discretizes spacetime and numerically solves QCD on a finite grid. Lattice QCD has made significant progress in recent years, providing first-principles calculations of nucleon FFs and radii with increasingly precise results~\cite{CSSM:2014knt,Shanahan:2014cga,QCDSF:2017ssq,Hasan:2017wwt,Park:2021ypf,Alexandrou:2017ypw,Alexandrou:2018sjm,Alexandrou:2020aja,Shintani:2018ozy,Ishikawa:2021eut,Tsuji:2023llh,Jang:2019jkn,Djukanovic:2021cgp,Djukanovic:2023jag,Djukanovic:2023beb,Syritsyn:2023pmn}. Another important framework is holographic QCD which is a theoretical framework that uses the anti de Sitter space/conformal field theory (AdS/CFT) correspondence to study QCD in the non-perturbative regime~\cite{Abidin:2009hr,Brodsky:2014yha,Sufian:2016hwn,Gutsche:2019jzh,Contreras:2021epz,Mamo:2021jhj,Ahmady:2021qed,Mondal:2019jdg,Xu:2021wwj}. Additionally, Dyson-Schwinger equations (DSEs) and Bethe-Salpeter approaches offer a continuum formulation of QCD, enabling the calculation of form factors by solving integral equations for quark and gluon propagators~\cite{Maris:2003vk,Eichmann:2009qa,Eichmann:2016yit,Burkert:2017djo,Ding:2022ows,Ferreira:2023fva,Barabanov:2020jvn,Yao:2024uej}. It is worth pointing in this context to various chiral theories and models~\cite{Gasser:1987rb,Kubis:2000aa,Alarcon:2012nr,Bauer:2012pv,Liu:2013fda,Flores-Mendieta:2015wir,He:2017viu,HillerBlin:2017syu,Yang:2020rpi,Bar:2021crj,Alvarado:2023loi,Aung:2024qmf}. For example, $ \chi $PT exploits the symmetries of QCD in the low-energy regime and provides systematic expansions in terms of the pion mass and momentum. $ \chi $PT has been particularly useful for understanding the pion cloud contributions to nucleon FFs and radii.  
There are also other approaches such as the constituent quark models, which describe nucleons as bound states of three quarks interacting via phenomenological potentials~\cite{Berger:2004yi,Julia-Diaz:2003sdv,Sharma:2013mfa,Plessas:2015mpa} and LCSR, which is a theoretical framework used to study the properties of hadrons by combining the principles of QCD sum rules with the light-cone distribution amplitudes of hadrons~\cite{Balitsky:1985ag,Lenz:2003tq,Castillo:2003pt,Braun:2006hz,Aliev:2008cs,Anikin:2013aka}. These theoretical frameworks, while differing in their assumptions and methodologies, collectively provide valuable insights into the electromagnetic structure of the nucleon and help bridge the gap between experimental data and fundamental QCD principles.

\begin{figure}[t!]
\scalebox{0.7}{\input{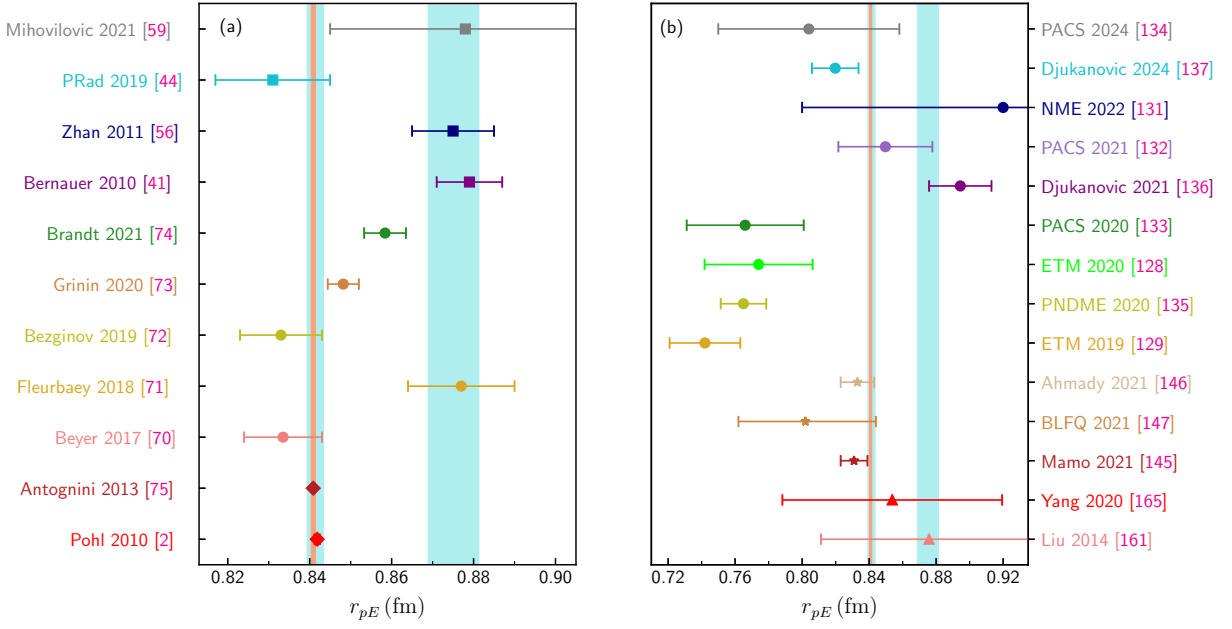}}
\caption{A comparison of the proton charge radius \( r_{pE} \) obtained from different (a) experimental techniques and (b) theoretical approaches. The results highlight the discrepancies and agreements between various methods, including spectroscopy, scattering experiments, lattice QCD, Holographic QCD, and chiral calculations. See text for more information.}
\label{Fig:fig1}
\end{figure}

Figure~\ref{Fig:fig1} (left panel) shows a comparison between the results obtained for the proton charge radius $ r_{pE} $ from different experimental techniques. In this figure (as well as the next figure), the right band and the left wider band correspond to the CODATA 2014~\cite{Mohr:2015ccw} and CODATA 2018~\cite{Tiesinga:2021myr} values, respectively, while the left narrow band represents the PDG 2024~\cite{ParticleDataGroup:2024cfk} value. For studies where multiple sources of uncertainties were provided, we have added them in quadrature to obtain a single combined error. As can be seen, the results clearly illustrate the proton charge radius puzzle. While muonic hydrogen spectroscopy (diamonds) consistently yields smaller values of $ r_{pE} $~\cite{Pohl:2010zza,Antognini:2013txn}, the results from ordinary hydrogen spectroscopy (circles) span a wide range. Notably, two recent measurements from hydrogen spectroscopy~\cite{Grinin:2020txk,Brandt:2021yor} have reported moderate values for $ r_{pE} $. Among the scattering experiments~\cite{Xiong:2019umf,A1:2010nsl,Zhan:2011ji,Mihovilovic:2019jiz} (squares), only the PRad experiment~\cite{Xiong:2019umf} is in agreement with the muonic hydrogen spectroscopy results. It is also worth noting that muonic hydrogen spectroscopy achieves significantly smaller uncertainties compared to both scattering experiments and ordinary hydrogen spectroscopy. Overall, these experimental results cover the interval $ 0.831 < r_{pE} < 0.879 $ fm. A simple combination of the scattering experiments yields $ r_{pE} = 0.866 \pm 0.010 $ fm, while the corresponding value from hydrogen spectroscopy is $ r_{pE} = 0.848 \pm 0.003 $ fm, which is approximately 1.85\% smaller than the average of the scattering experiment results. The simple combination of all experimental results leads to $ r_{pE} = 0.857 \pm 0.005 $ fm. We note that these combined values represent unweighted averages of independent results, with uncertainties propagated through standard error analysis assuming no correlations. This approach is adopted throughout this article to calculate the combined values. 

In the right panel of Fig.~\ref{Fig:fig1}, we compare the results obtained for \( r_{pE} \) from various theoretical approaches. The primary results come from lattice QCD~\cite{Park:2021ypf,Alexandrou:2018sjm,Alexandrou:2020aja,Shintani:2018ozy,Ishikawa:2021eut,Tsuji:2023llh,Jang:2019jkn,Djukanovic:2021cgp,Djukanovic:2023jag} (circles). Additionally, we include results from holographic QCD~\cite{Mamo:2021jhj,Ahmady:2021qed,Xu:2021wwj} (stars) and chiral calculations~\cite{Liu:2013fda,Yang:2020rpi} (triangles). As can be seen, most theoretical results exhibit significant uncertainties. The results clearly show that lattice QCD predictions have not yet converged. This variability complicates direct comparison with experimental determinations and highlights the need for continued improvements in lattice methodologies. For instance, although the NME Collaboration~\cite{Park:2021ypf} obtained a relatively large value, its uncertainty is also substantial, with the lower band even overlapping the lower range of all experimental results. Overall, the theoretical results span the intervals \( 0.742 < r_{pE} < 0.92 \) fm (including lattice results) and \( 0.802 < r_{pE} < 0.878 \) fm (excluding lattice results). The corresponding simple combinations are \( r_{pE} = 0.827 \pm 0.013 \) fm and \( r_{pE} = 0.839 \pm 0.020 \) fm, respectively. The simple combination of the lattice results alone yields \( r_{pE} = 0.815 \pm 0.016 \) fm, indicating that lattice QCD predictions tend to yield smaller values for \( r_{pE} \) compared to other theoretical approaches and experimental results. Moreover, it is evident that theoretical approaches generally predict smaller values for \( r_{pE} \) compared to experimental measurements.

\begin{figure}[t!]
\scalebox{0.7}{\input{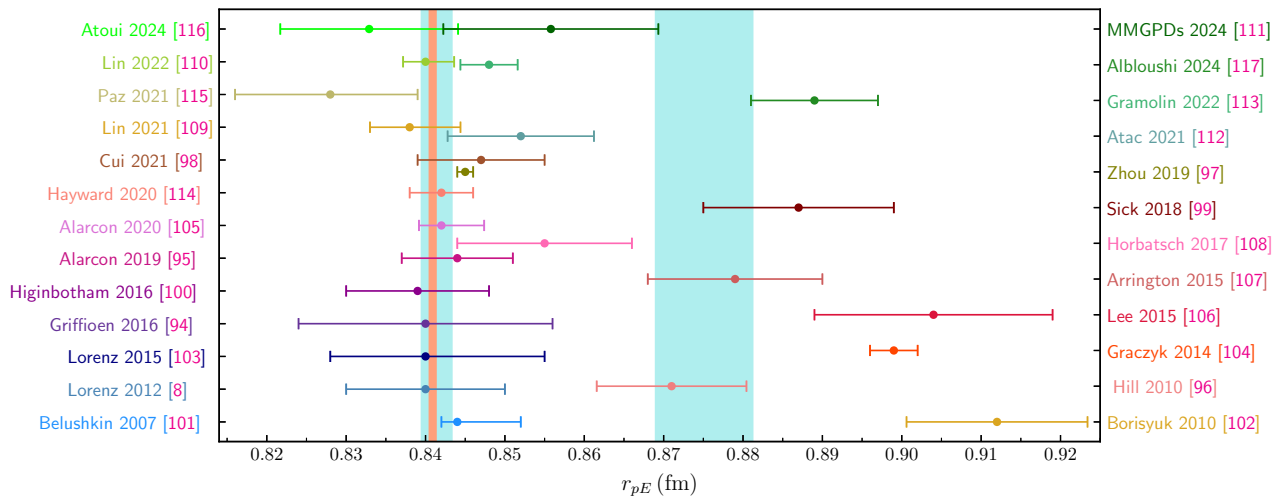}}
\caption{A comparison between the results obtained for the proton charge radius \( r_{pE} \) from different analyses and phenomenological studies.}
\label{Fig:fig2}
\end{figure}

Figure~\ref{Fig:fig2} shows a comparison between the results obtained for the proton charge radius \( r_{pE} \) from different analyses and phenomenological studies~\cite{Hill:2010yb,Lorenz:2012tm,Lorenz:2014yda,Griffioen:2015hta,Alarcon:2018zbz,Alarcon:2020kcz,Cui:2021vgm,Lee:2015jqa,Sick:2018fzn,Arrington:2015ria,Horbatsch:2016ilr,Higinbotham:2015rja,Lin:2021umk,Lin:2021xrc,Zhou:2018bon,Atac:2020hdq,Gramolin:2021gln,Hayward:2018qij,Paz:2020prs,Borisyuk:2009mg,Graczyk:2014lba,Atoui:2023nrn,Albloushi:2024wxc,Belushkin:2006qa,Goharipour:2024mbk}. Note that for Ref.~\cite{Lee:2015jqa}, we have taken the value obtained by a simple combination of the Mainz and world values. Overall, the results are predominantly located in the region corresponding to smaller values of \( r_{pE} \). The interval covered by these studies is \( 0.828 < r_{pE} < 0.912 \) fm. A simple combination of all results leads to \( r_{pE} = 0.857 \pm 0.002 \) fm. Interestingly, this value is in excellent agreement with the simple combination of all experimental results, which is \( r_{pE} = 0.857 \pm 0.005 \) fm. This is also consistent with the recent findings of the MMGPDs Collaboration~\cite{Goharipour:2024mbk}, which demonstrated that analyzing all available electron-nucleon scattering data simultaneously within the framework of GPDs yields an average value of \( r_{pE} = 0.856 \pm 0.013 \) fm, albeit with a relatively large uncertainty. It is also insightful to combine all results obtained from experimental, theoretical, and phenomenological efforts. Doing so, we obtain \( r_{pE} = 0.847 \pm 0.005 \) fm.

\begin{figure}[t!]
\scalebox{0.7}{\input{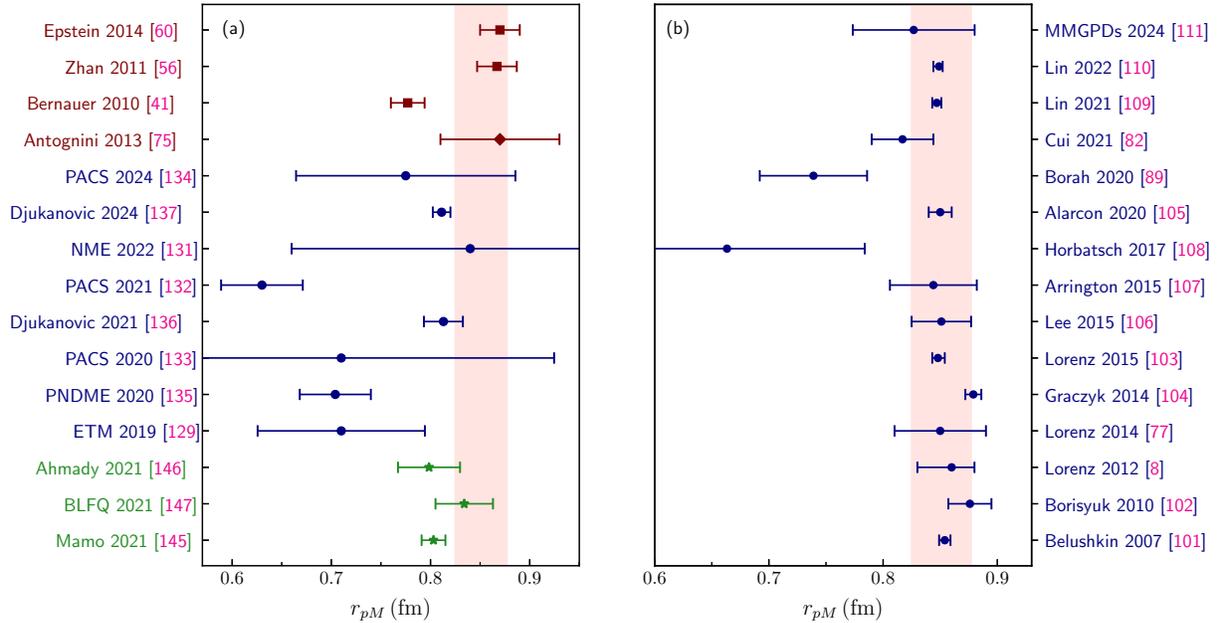}}
\caption{A comparison of the proton magnetic radius \( r_{pM} \) obtained from (a) different experimental techniques (squares) and theoretical approaches, including lattice QCD (circles) and holographic QCD (stars), and (b) various analyses and phenomenological studies. The results highlight the agreements and discrepancies between different methods, providing insights into the proton's magnetic structure. See text for more details.}
\label{Fig:fig3}
\end{figure}

Now we turn our attention to the proton magnetic radius \( r_{pM} \). Compared to the charge radius, there is less information available for \( r_{pM} \). In Fig.~\ref{Fig:fig3} (left panel), we compare the available experimental results, including the scattering experiments (squares)~\cite{A1:2010nsl,Zhan:2011ji,Epstein:2014zua} and muonic hydrogen spectroscopy (diamonds)~\cite{Antognini:2013txn}, as well as the theoretical results, including lattice QCD (circles)~\cite{Park:2021ypf,Alexandrou:2018sjm,Shintani:2018ozy,Ishikawa:2021eut,Tsuji:2023llh,Jang:2019jkn,Djukanovic:2021cgp,Djukanovic:2023jag} and holographic QCD (stars)~\cite{Mamo:2021jhj,Ahmady:2021qed,Xu:2021wwj}, alongside the PDG 2024 value~\cite{ParticleDataGroup:2024cfk}, which is represented by a band. As can be seen, the agreement between different results is better compared to the case of the proton charge radius. The smallest values of \( r_{pM} \) are again obtained from lattice QCD calculations. Considering both theory and experiment, \( r_{pM} \) lies in the interval \( 0.630 < r_{pM} < 0.87 \) fm. A simple combination of all experimental results yields \( r_{pM} = 0.846 \pm 0.017 \) fm, while the combination of all lattice predictions gives \( r_{pM} = 0.759 \pm 0.040 \) fm, and the combination of all theoretical results yields \( r_{pM} = 0.786 \pm 0.021 \) fm. Notably, in all cases, the combined value for \( r_{pM} \) is smaller than that for \( r_{pE} \), with larger uncertainties. While the difference between the combined theoretical results for \( r_{pE} \) and \( r_{pM} \) is significant (approximately 0.04 fm), it is just about 0.01 fm for the combined experimental results. 

In the right panel, we compare the results obtained from different analyses and phenomenological studies~\cite{Lorenz:2012tm,Lorenz:2014vha,Lorenz:2014yda,Alarcon:2020kcz,Lee:2015jqa,Arrington:2015ria,Horbatsch:2016ilr,Lin:2021umk,Lin:2021xrc,Cui:2021skn,Borah:2020gte,Borisyuk:2009mg,Graczyk:2014lba,Belushkin:2006qa,Goharipour:2024mbk}. These results cover the interval \( 0.663 < r_{pM} < 0.879 \) fm, with a simple combination of \( r_{pM} = 0.830 \pm 0.011 \) fm. This value is smaller than the combined experimental results but larger than the combined theoretical results. It is also worth noting that the difference between the combined experimental and phenomenological results (0.016 fm) is more significant compared to the case of \( r_{pE} \). Finally, combining all experimental, theoretical, and phenomenological results, we obtain \( r_{pM} = 0.821 \pm 0.01 \) fm.

As mentioned before, the determination of the neutron's charge and magnetic radii is more challenging compared to the proton due to several intrinsic properties and experimental limitations. For example, since the neutron cannot be isolated as a free target for scattering experiments, its properties must be inferred from experiments involving deuterium or other light nuclei. This introduces additional complexities, as the measurements must account for the nuclear corrections. Moreover, free neutrons have a short lifetime (about 15 minutes), decaying into a proton, electron, and antineutrino. This limits the time available for precise measurements and requires specialized experimental setups to work with neutron beams or targets. Note also that the neutron is electrically neutral. This neutrality makes it difficult to use electromagnetic probes (such as electron scattering) to directly measure the neutron's charge distribution. As a result, less information is available on the neutron's charge and magnetic radii compared with the proton. 

Figure~\ref{Fig:fig4} (left panel) shows a comparison between the results obtained for the mean-square charge radius of the neutron \( \left<r_{nE}^2\right> \) from different experimental techniques (squares)~\cite{Krohn:1973re,Koester:1995nx,Kopecky:1997rw,Atac:2021wqj,Heacock:2021btd}, as described earlier, as well as the corresponding results from various analyses and phenomenological studies (circles)~\cite{Atac:2020hdq,Filin:2019eoe,Filin:2020tcs,Albloushi:2024wxc,Belushkin:2006qa,Goharipour:2024mbk}. The PDG value~\cite{ParticleDataGroup:2024cfk} is represented by a vertical band. As can be seen, the results are in better agreement compared to the proton case. In the right panel, we compare the results obtained from different theoretical techniques, including lattice QCD (circles)~\cite{Alexandrou:2018sjm,Alexandrou:2020aja,Shintani:2018ozy,Djukanovic:2023jag,Tsuji:2023llh}, holographic QCD (stars)~\cite{Mamo:2021jhj,Xu:2021wwj,Ahmady:2021qed}, and chiral methods (up triangles)~\cite{Liu:2013fda,Yang:2020rpi}. Note that Trinhammer~\cite{Trinhammer:2021irt} obtained \( \left<r_{nE}^2\right> \) by assuming the neutron structure to be of intrinsic origin, determined by a mass Hamiltonian on the intrinsic configuration space \( U(3) \). Overall, the experimental, phenomenological, and theoretical results span the intervals \( -0.124 < \left<r_{nE}^2\right> < -0.110 \) fm\(^2\), \( -0.122 < \left<r_{nE}^2\right> < -0.089 \) fm\(^2\), and \( -0.146 < \left<r_{nE}^2\right> < -0.014 \) fm\(^2\), respectively. Their corresponding combined values are \( \left<r_{nE}^2\right> = -0.115 \pm 0.002 \) fm\(^2\), \( \left<r_{nE}^2\right> = -0.110 \pm 0.005 \) fm\(^2\), and \( \left<r_{nE}^2\right> = -0.076 \pm 0.020 \) fm\(^2\). Finally, combining all experimental, theoretical, and phenomenological results, we obtain \( \left<r_{nE}^2\right> = -0.100 \pm 0.007 \) fm\(^2\).
\begin{figure}[t!]
\scalebox{0.7}{\input{rEn2.pgf}}
\caption{A comparison of the mean-square charge radius of the neutron \( \left<r_{nE}^2\right> \) obtained from (a) different experimental techniques and various analyses and phenomenological studies, and (b) different theoretical approaches. The results highlight the discrepancies and agreements between various methods. The vertical band shows the PDG value~\cite{ParticleDataGroup:2024cfk}. See text for more information.}
\label{Fig:fig4}
\end{figure}

\begin{figure}[t!]
\scalebox{0.7}{\input{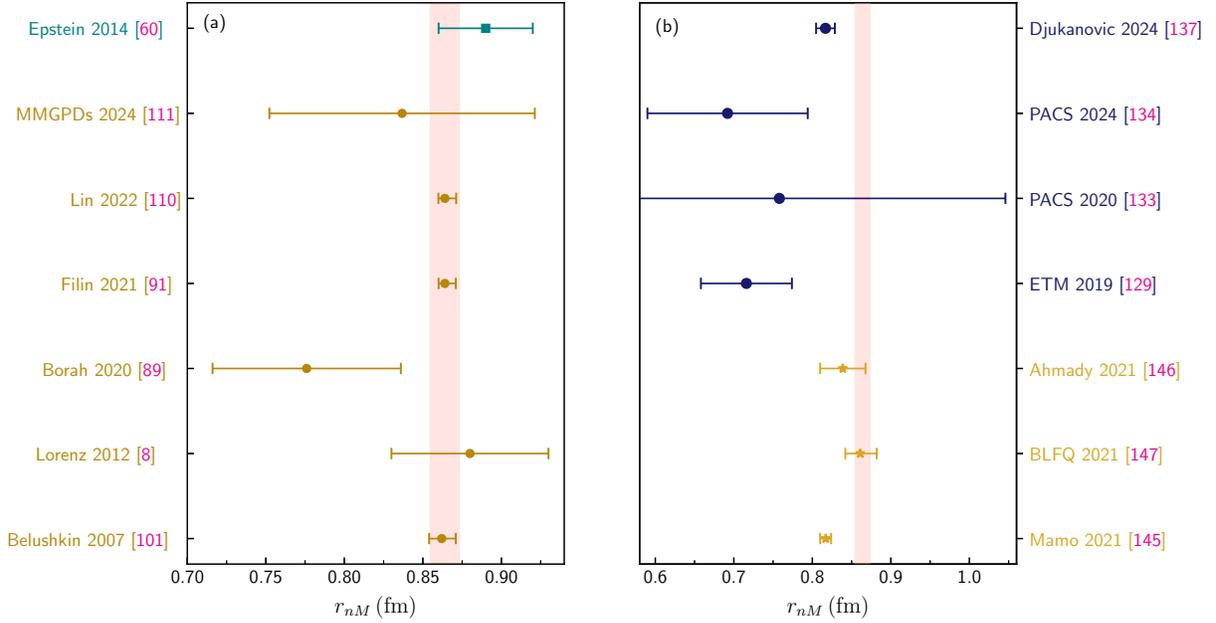}}
\caption{A comparison of the magnetic radius of the neutron \( r_{nM} \) obtained from (a) different experimental techniques and various analyses and phenomenological studies, and (b) different theoretical approaches. The results highlight the discrepancies and agreements between various methods. The vertical band shows the PDG value~\cite{ParticleDataGroup:2024cfk}. See text for more information.}
\label{Fig:fig5}
\end{figure}

Figure~\ref{Fig:fig5} presents results for the neutron magnetic radius \( r_{nM} \), analogous to those shown in Fig.~\ref{Fig:fig4}. Notably, there is a single experimental measurement by Epstein et al.~\cite{Epstein:2014zua}, reporting \( r_{nM} = 0.890 \pm 0.030 \) fm. Theoretical predictions, especially from lattice QCD, generally indicate smaller values. The entire set of experimental, phenomenological, and theoretical results spans the range  \( 0.692 < r_{nM} < 0.890 \) fm, with a combined value of \( r_{nM} = 0.841 \pm 0.019 \) fm. The combined value of the phenomenological and theoretical studies is \( r_{nM} = 0.847 \pm 0.019 \) fm and \( r_{nM} = 0.786 \pm 0.045 \) fm, respectively. Note that the later is exactly same as the proton case, \( r_{pM} = 0.786 \pm 0.021 \) fm, but with larger uncertainty.

Each of the above experimental, phenomenological, and theoretical studies has contributed to our understanding of the nucleon's charge and magnetic radii, but they also highlight the challenges in achieving consistent results across different methods. Discrepancies, such as the proton radius puzzle, underscore the importance of cross-checking measurements and refining both experimental techniques and theoretical frameworks to achieve a unified understanding of nucleon structure. Key ongoing and upcoming experiments include the MUSE experiment at PSI, which compares elastic scattering cross sections using muon and electron beams to test lepton universality; the COMPASS++/AMBER experiment at CERN, which measures high-energy muon-proton scattering at low \( Q^2 \); PRad-II experiment at Jefferson Lab, which aims to significantly improve precision in electron-proton scattering measurements through advanced detector upgrades and reduced systematic uncertainties, and  DRad Experiment at Jefferson Lab, which measures the deuteron charge radius ($ r_d $) using a high-precision elastic electron-deuteron scattering experiment. Additionally, future experiments at Mainz (PRES and MAGIX) and Tohoku University (UL\({\rm Q}^2\)) focus on low \( Q^2 \) measurements using innovative techniques such as recoil proton detection and windowless gas-jet targets. These experiments collectively aim to achieve unprecedented precision, with PRad-II targeting an uncertainty of \(\sim 0.0036\) fm, while also testing for potential lepton universality violations and refining our understanding of the proton's structure. By employing diverse methodologies and cutting-edge technologies, these efforts seek to resolve the proton radius puzzle and provide deeper insights into fundamental physics. For a more detailed review, see Ref.~\cite{Gao:2021sml,Gao:2024gvk} and references therein.


\section{The axial charge radius}\label{sec:three}
%
%

As mentioned in the Introduction, the axial radius is associated with the distribution of the nucleon's weak charge, as probed in weak interaction processes such as beta decay or neutrino scattering. It reflects the spatial extent of the nucleon's involvement in weak interactions and provides a measure of the nucleon's "size" in the context of weak forces. However, it is not directly comparable to the charge or magnetic radii, as it probes a fundamentally different aspect of the nucleon's structure. Similar to the case of the charge and magnetic radii, the proton's mean-squared axial charge radius \( \left<r_A^2\right> \) (and hence the axial charge radius $ r_A $) can be calculated from the axial form factor (AFF) \( G_A(t) \) (often denoted as \( F_A \) in the literature) using the following relation~\cite{MINERvA:2023avz}:
\begin{equation}
\label{Eq2}
\langle r_A^2\rangle = \left. \frac{6}{G_A(0)} \dv{G_A}{t}\right|_{t=0},
\end{equation}
where \( g_A = G_A(0) \) is commonly referred to as the axial charge or axial coupling constant.

From past to present, there have been significant efforts to determine \( G_A(t) \), \( g_A \), and \( r_A \) as well as the dipole mass $ M_A $ using experimental techniques~\cite{Bernard:2001rs,MINERvA:2023avz,Ahrens:1986xe,K2K:2006odf,CLAS:2012ich,T2K:2014hih,Petti:2023abz,NOMAD:2009qmu,MiniBooNE:2010bsu,MiniBooNE:2010xqw,Butkevich:2013vva}, data analyses and phenomenological studies~\cite{Adamuscin:2007fk,Bodek:2007ym,Meyer:2016oeg,Hill:2017wgb,Alvarez-Ruso:2018rdx,Unal:2018ruo,Kaiser:2024vbc,Irani:2023lol,Hashamipour:2019pgy,Tomalak:2020zlv,Megias:2019qdv,Pate:2024acz}, and theoretical calculations~\cite{Mondal:2019jdg,Xu:2021wwj,Sasaki:2007gw,Yamazaki:2009zq,LHPC:2010jcs,Capitani:2012gj,Horsley:2013ayv,Bali:2014nma,Green:2012ud,Hasan:2017wwt,Bhattacharya:2016zcn,Alexandrou:2017hac,Alexandrou:2020okk,Alexandrou:2021wzv,Alexandrou:2023qbg,Park:2021ypf,Ishikawa:2018rew,Shintani:2018ozy,Ishikawa:2021eut,Tsuji:2023llh,Gupta:2017dwj,Jang:2023zts,Djukanovic:2022wru,RQCD:2019jai,Capitani:2017qpc,Green:2017keo,Jang:2019vkm,Chang:2018uxx,Alexandrou:2024ozj,Meyer:2022mix,Bhattacharya:2024wtg,Tomalak:2023pdi,Yao:2017fym,Lutz:2020dfi,Hermsen:2024eth,Atayev:2022omk,Mamedov:2022yyd,Schindler:2006it,Flores-Mendieta:2006ojy,Liu:2018jiu,Liu:2022ekr,Dahiya:2014jfa,Wang:2021ild,Eichmann:2011pv,Ramalho:2015jem,Choi:2010ty,Ramalho:2024tdi,Chen:2020wuq,Chen:2021guo,Braun:2006hz,Wang:2006su,Aliev:2007qu,Anikin:2016teg,Chen:2024oxx,Chen:2024ksq,Hermsen:2025vds}. From an experimental perspective, our knowledge of the axial form factors (AFFs) is more limited compared to the electromagnetic FFs. The primary sources of information come from two classes of experiments: (anti)neutrino scattering off protons or nuclei and charged pion electroproduction. Additionally, muon capture by a proton and parity-violating electron scattering provide alternative methods to probe the AFF at very low and higher momentum transfers, respectively. For neutrino scattering experiments, notable measurements include: 
the measurement of the double differential cross section for charged-current quasielastic (CCQE) scattering on carbon and the corresponding AFF by the MiniBooNE Collaboration~\cite{MiniBooNE:2010bsu,Butkevich:2013vva} and the first high-statistics measurement of the \( \bar{\nu}_\mu p \rightarrow \mu^{+} n \) reaction by the MINERvA Collaboration~\cite{MINERvA:2023avz}. For charged pion electroproduction, numerous measurements have been conducted over the decades. A particularly important and recent example is the measurements performed by the CLAS Collaboration~\cite{CLAS:2012ich} through the extraction of the pion-nucleon multipoles near the production threshold for the \( n \pi^{+} \) channel at relatively high momentum transfer. For a comprehensive review and discussion of these topics, see Refs.~\cite{Bernard:2001rs,Hashamipour:2019pgy} and references therein.
 
For the case of data analyses and phenomenological studies, one can refer to the following works. In Ref.~\cite{Bodek:2007ym}, Bodek \textit{et al.} obtained parameterizations of both vector and axial nucleon FFs through the analysis of experimental data from neutrino-deuteron scattering and pion electroproduction experiments. Meyer \textit{et al.}~\cite{Meyer:2016oeg} performed an analysis of world data for neutrino-deuteron scattering using a model-independent and systematically improvable parameterization of \( G_A \). Hill \textit{et al.}~\cite{Hill:2017wgb} used muon capture measurements of the singlet muonic hydrogen capture rate to determine \( r_A \). They indicated that its uncertainty can be reduced by combining muon capture data with the independent \( z \)-expansion neutrino scattering results from Ref.~\cite{Meyer:2016oeg}. In Ref.~\cite{Alvarez-Ruso:2018rdx}, Alvarez-Ruso \textit{et al.} applied a Bayesian neural-network analysis of neutrino-scattering data to extract AFF \( G_A \) and, consequently, \( r_A \) in a model-independent way. In contrast, \"{U}nal \textit{et al.}~\cite{Unal:2018ruo} determined these quantities by analyzing the low-\( Q^2 \) behavior of the AFF within a chiral effective-Lagrangian model that includes the \( a_1 \) meson, fitting the relevant coupling parameters to experimental data. Irani \textit{et al.}~\cite{Irani:2023lol} predicted \( r_A \) by performing a QCD analysis of the MINERvA data, combined with all available data on the proton's AFF, using the framework of GPDs. Finally, Kaiser \textit{et al.}~\cite{Kaiser:2024vbc} presented a phenomenological determination of nucleon properties, including the AFF and charge radius, using a spectral analysis to separate contributions to the nucleon's form factors into low-mass (mesonic) and high-mass (short-range) components.

The theoretical calculations of the axial properties of the nucleon, including \( G_A(t) \), \( g_A \), \( r_A \), and \( M_A \), have been predominantly performed within the lattice QCD framework~\cite{Alexandrou:2017hac,Alexandrou:2020okk,Alexandrou:2021wzv,Alexandrou:2023qbg,Park:2021ypf,Ishikawa:2018rew,Shintani:2018ozy,Ishikawa:2021eut,Tsuji:2023llh,Gupta:2017dwj,Jang:2023zts,Djukanovic:2022wru,RQCD:2019jai,Capitani:2017qpc,Green:2017keo,Jang:2019vkm,Yamazaki:2009zq,LHPC:2010jcs,Capitani:2012gj,Horsley:2013ayv,Bali:2014nma,Green:2012ud,Bhattacharya:2016zcn,Hasan:2017wwt,Sasaki:2007gw,Chang:2018uxx,Alexandrou:2024ozj,Meyer:2022mix,Bhattacharya:2024wtg,Tomalak:2023pdi} (see Ref.~\cite{Gupta:2024qip} for an overview). However, significant investigations have also been conducted using other theoretical techniques, including holographic QCD~\cite{Mondal:2019jdg,Xu:2021wwj,Atayev:2022omk,Mamedov:2022yyd}, chiral methods such as \(\chi\)PT~\cite{Yao:2017fym,Lutz:2020dfi,Schindler:2006it,Flores-Mendieta:2006ojy}, the perturbative chiral quark model~\cite{Liu:2018jiu,Liu:2022ekr}, and the chiral constituent quark model~\cite{Dahiya:2014jfa,Wang:2021ild}. Additionally, studies have employed DSEs approach~\cite{Eichmann:2011pv}, covariant spectator quark model~\cite{Ramalho:2015jem}, relativistic constituent quark model~\cite{Choi:2010ty,Ramalho:2024tdi}, quark+diquark model~\cite{Chen:2020wuq,Chen:2021guo}, and LCSR~\cite{Braun:2006hz,Wang:2006su,Aliev:2007qu,Anikin:2016teg}. Recently, Chen \textit{et al.}~\cite{Chen:2024oxx} demonstrated that the slope of \( G_A \) in the forward limit, as given in Eq.~(\ref{Eq2}), does not represent the three-dimensional mean-square axial radius in the Breit frame but instead corresponds to a contribution to the mean-square spin radius.

\begin{figure}[t!]
\scalebox{0.7}{\input{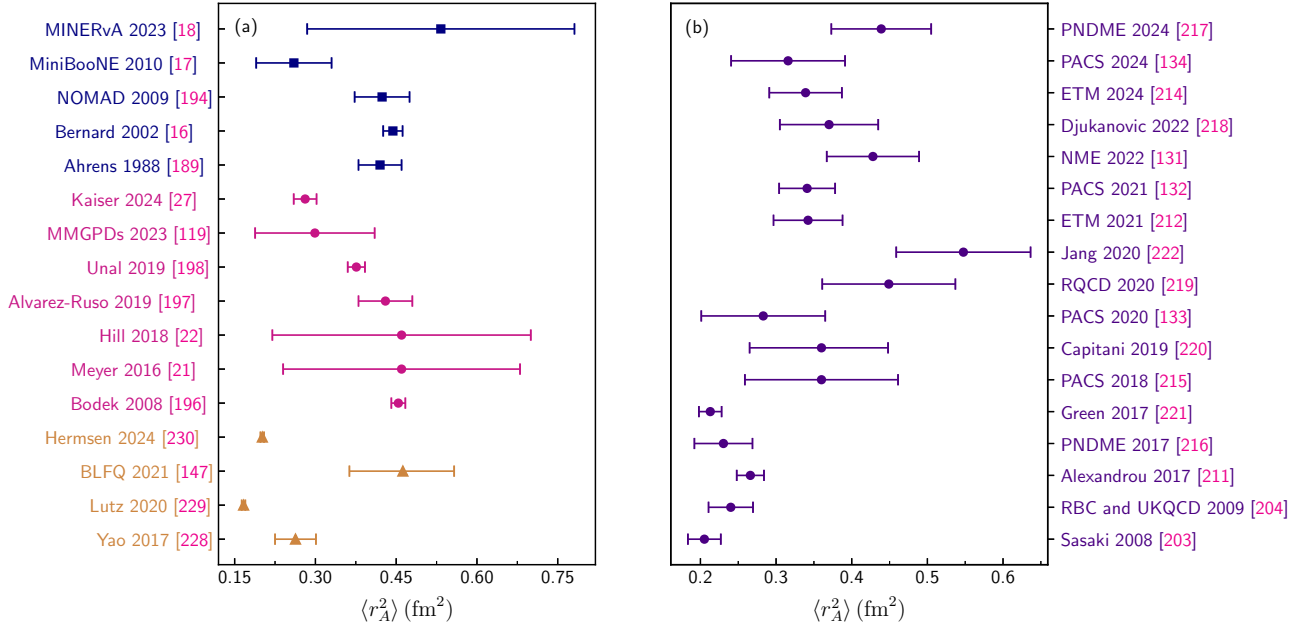}}
\caption{A comparison of the mean-squared axial charge radius \( \left<r_A^2\right> \) of the proton obtained from (a) different experimental techniques, various data analyses and phenomenological studies, and theoretical approaches, and (b) lattice QCD calculations. The results highlight the discrepancies and agreements among various methods. See text for more information.}
\label{Fig:fig6}
\end{figure}

Figure~\ref{Fig:fig6} (left panel) shows a comparison between the results obtained for the mean-square axial charge radius of the proton \( \left<r_A^2\right> \) from different experimental techniques (squares)~\cite{MINERvA:2023avz,Ahrens:1986xe,MiniBooNE:2010bsu,NOMAD:2009qmu,Bernard:2001rs}, various data analyses and phenomenological studies (circles)~\cite{Hill:2017wgb,Meyer:2016oeg,Bodek:2007ym,Alvarez-Ruso:2018rdx,Unal:2018ruo,Irani:2023lol,Kaiser:2024vbc}, and different theoretical approaches (up triangles), including holographic QCD and \(\chi\)PT. The lattice QCD results are compared separately in the right panel. Note that for the experimental results, we have used the relation \( \left<r_A^2\right> = 12/M^2_A \) with the extracted value of \( M_A \) to calculate \( \left<r_A^2\right> \) wherever direct values were not provided. As can be seen from the figure, the results from different studies span a wide range of \( \left<r_A^2\right> \) from 0.167 to 0.548 fm\(^2\). A simple combination of the experimental, phenomenological, lattice QCD, and all theoretical results yields \( r_A= 0.645 \pm 0.101 \) fm, \( r_A= 0.628 \pm 0.093 \) fm, \( r_A= 0.573 \pm 0.051 \) fm, and \( r_A= 0.533^{+0.071}_{-0.072} \) fm, respectively. While the combined values of experiment and phenomenology exhibit good consistency, the combined value of the theoretical studies shows a significant discrepancy. Combining all experimental, phenomenological, and theoretical results, we obtain \( r_A= 0.604 \pm 0.053 \) fm, which is notably smaller compared to the corresponding values obtained for the charge and magnetic radii of the proton in the previous section.
%


\section{The mechanical and mass radii}\label{sec:four} 

As mentioned in the Introduction, the mechanical and mass radii provide critical insights into the spatial distribution of the nucleon's internal energy and pressure~\cite{Polyakov:2018zvc,Lorce:2018egm,Burkert:2023wzr}, i.e., the mechanical properties of the nucleon. Specifically, the mechanical radius, derived from EMT, describes the spatial distribution of pressure and shear forces within the nucleon, while the mass radius reflects the spatial distribution of the nucleon's mass, which is primarily generated by the energy of the strong interaction. Theoretically, these radii can be calculated from the gravitational FFs (GFFs) \( D(t) \) and \( A(t) \) of the EMT as follows~\cite{Polyakov:2018zvc}:
\begin{align}
    \langle r^2 \rangle_{\text{mech}} &= \frac{6D(0)}{\int_{-\infty}^0 D(t)\, dt} \,, \label{Eq3} \\
    \langle r^2 \rangle_{\text{mass}} &= 6 A^{\prime}(0) -\frac{3D(0)}{2m^2} \,, \label{Eq4}
\end{align}
where \( m \) is the nucleon mass (note that the mechanical radius depends solely on \( D(t) \)). Conceptually, \( D(t) \) can be interpreted as a measure of the forces that bind the hadron together, particularly in terms of the pressure and shear forces exerted by quarks and gluons, while \( A(t) \) provides information on the fractions of the momentum carried by the quark and gluon constituents of the nucleon.

From past to present, there have been significant efforts to study the mechanical properties of the nucleon~\cite{Lorce:2018egm,Pagels:1966zza,Lorce:2015lna,Teryaev:2016edw,Lowdon:2017idv,Cotogno:2019xcl,Varma:2020crx,Ji:2021mtz,Ji:2021mfb,Freese:2021czn,Cao:2023ohj,Won:2023cyd,Yao:2024ixu,Li:2024kas,Burkert:2018bqq,Kumericki:2019ddg,Burkert:2023atx,Goharipour:2025lep,Mamo:2019mka,Kharzeev:2021qkd,Duran:2022xag,Guo:2021ibg,Guo:2023pqw,Guo:2023qgu,Han:2022qet,Wang:2019mza,Wang:2021dis,Wang:2021ujy,Wang:2022vhr,Wang:2022uch,Wang:2022ndz,Wang:2022zwz,Wang:2023fmx,Hatta:2025vhs,Mamo:2022eui,Belitsky:2005qn,Polyakov:2002yz,Shanahan:2018nnv,Shanahan:2018pib,Pefkou:2021fni,Hackett:2023rif,Wang:2024lrm,Mamo:2021krl,deTeramond:2021lxc,Fujita:2022jus,Fujii:2024rqd,Sugimoto:2025btn,Anikin:2019kwi,Azizi:2019ytx,Dehghan:2025ncw,Jung:2013bya,Jung:2014jja,Goeke:2007fp,Kim:2020nug,Kim:2021jjf,Won:2022cyy,Won:2023zmf,Won:2023ial,Cebulla:2007ei,Kim:2012ts,GarciaMartin-Caro:2023klo,Fujii:2025aip,Neubelt:2019sou,Owa:2021hnj,Neubelt:2019sou,Owa:2021hnj,Chakrabarti:2020kdc,Choudhary:2022den,More:2021stk,Nair:2024fit,Cao:2024zlf,Kou:2023azd,Broniowski:2025ctl,Sain:2025kup}, including GFFs, internal forces, pressure, surface tension, mechanical and mass radii, geometric shape, and more. These studies have utilized a wide range of theoretical approaches and phenomenological frameworks. For phenomenological studies, one can refer to analyses based on experimental data from deeply virtual Compton scattering (DVCS)~\cite{Burkert:2018bqq,Kumericki:2019ddg,Burkert:2023atx,Goharipour:2025lep} and near-threshold photo-production of \( J/\Psi \), \( \omega \), and \( \phi \) particles~\cite{Mamo:2019mka,Kharzeev:2021qkd,Duran:2022xag,Guo:2021ibg,Guo:2023pqw,Guo:2023qgu,Han:2022qet,Wang:2019mza,Wang:2021dis,Wang:2021ujy,Wang:2022vhr,Wang:2022uch,Wang:2022ndz,Wang:2022zwz,Wang:2023fmx,Hatta:2025vhs}, as well as holographic fits to lattice data~\cite{Mamo:2022eui} and studies based on skewness-dependent GPDs~\cite{Belitsky:2005qn,Polyakov:2002yz}. In contrast, theoretical studies have employed a variety of models and approaches, including lattice QCD~\cite{Shanahan:2018nnv,Shanahan:2018pib,Pefkou:2021fni,Hackett:2023rif,Wang:2024lrm}, holographic QCD~\cite{Mamo:2021krl,deTeramond:2021lxc,Fujita:2022jus,Fujii:2024rqd,Sugimoto:2025btn}, LCSR~\cite{Anikin:2019kwi,Azizi:2019ytx,Dehghan:2025ncw}, various frameworks of soliton models~\cite{Jung:2013bya,Jung:2014jja,Goeke:2007fp,Kim:2020nug,Kim:2021jjf,Won:2022cyy,Won:2023zmf,Won:2023ial} (particularly the chiral quark-soliton model~\cite{Goeke:2007fp,Kim:2020nug,Kim:2021jjf,Won:2022cyy,Won:2023zmf,Won:2023ial}), the Skyrme model~\cite{Cebulla:2007ei,Kim:2012ts,GarciaMartin-Caro:2023klo,Fujii:2025aip}, the bag model~\cite{Neubelt:2019sou,Owa:2021hnj}, the light-front quark-diquark model~\cite{Chakrabarti:2020kdc,Choudhary:2022den}, the light-front Hamiltonian approach~\cite{More:2021stk,Nair:2024fit}, data-driven dispersive approaches~\cite{Cao:2024zlf}, configurational entropy methods~\cite{Kou:2023azd}, meson dominance model~\cite{Broniowski:2025ctl}, and light-front spectator model~\cite{Sain:2025kup}.

\begin{figure}[t!]
\scalebox{0.7}{\input{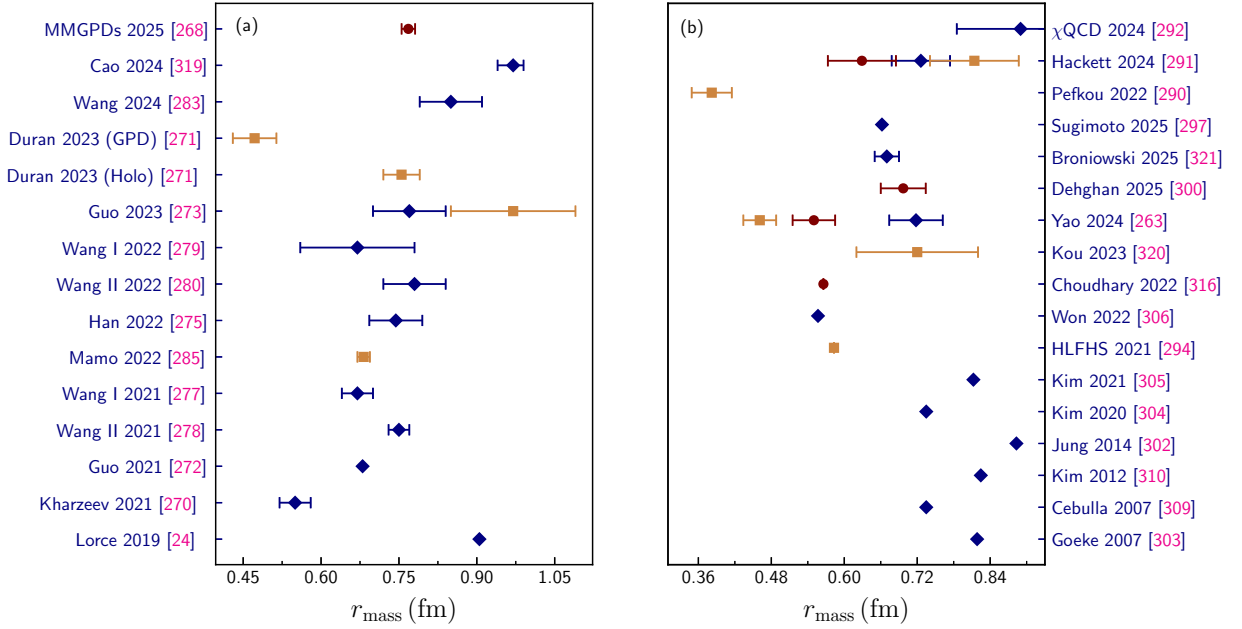}}
\caption{A comparison of the proton mass radius \( r_{\text{mass}} \) obtained from (a) various data analyses and phenomenological studies and (b) different theoretical techniques. The results highlight the discrepancies and agreements among various methods. Circle, square, and diamond symbols represent the quark, gluon, and total contributions, respectively. See text for more information.}
\label{Fig:fig7}
\end{figure}

Figure~\ref{Fig:fig7} shows a comparison between the results obtained for the proton mass radius \( r_{\text{mass}} \) from various data analyses and phenomenological studies (left panel) and different theoretical techniques (right panel). Several key points should be noted. First, different contributions to \( r_{\text{mass}} \), including the quark (circle), gluon (square), and total (diamond) contributions, are presented separately for each study. While most works calculated the total contribution, some studies focused solely on either the gluon or quark contributions. A few studies have provided results for all quark, gluon, and total contributions. Second, it is important to note that  GFFs are independent of the renormalization scale \( \mu \) when both quark and gluon (total) contributions are included. However, the quark and gluon contributions individually are \( \mu \)-dependent. Although most works present results at \( \mu = 2 \) GeV, some studies use different values of \( \mu \). For example, Burkert \textit{et al.}~\cite{Burkert:2023atx} performed their analysis at \( \mu^2 = 1.5 \) GeV\(^2\), LCSR calculations of Refs.~\cite{Azizi:2019ytx,Dehghan:2025ncw} were conducted at \( \mu = 1 \) GeV, and the result of the cloudy bag model by Owa \textit{et al.}~\cite{Owa:2021hnj} was obtained at \( \mu^2 = 1.4 \) GeV\(^2\). Third, in most studies, only the central value of \( r_{\text{mass}} \) has been reported without corresponding uncertainties. However, unlike in previous sections, we have chosen to include these results, given the limited number of extractions for the mechanical and mass radii of the nucleon. Finally, Han \textit{et al.}~\cite{Han:2022qet} obtained \( r_{\text{mass}} \) for a loosely bound proton rather than a free proton. Additionally, among the values obtained by Pefkou \textit{et al.}~\cite{Pefkou:2021fni}, we selected the result of the tripole BF3. As can be seen, the results from different studies span a wide range, particularly from \( 0.382 \) to \( 0.89 \) fm for theoretical calculations and from \( 0.472 \) to \( 0.970 \) fm for phenomenological studies. The corresponding combined values are \( r_{\text{mass}} = 0.727 \pm 0.025 \) fm and \( r_{\text{mass}} = 0.785 \pm 0.014 \) fm, respectively, indicating that theoretical calculations generally suggest smaller values for \( r_{\text{mass}} \) compared to phenomenological studies. Note also that, in the combination process, we have used the values of the total contributions for studies that presented also the quark or gluon contributions. Additionally, for studies that did not provide uncertainties, we have assumed the uncertainties to be zero. Combining all phenomenological and theoretical results, we obtain \( r_{\text{mass}}= 0.756 \pm 0.014 \) fm.

\begin{figure}[t!]
\scalebox{0.5}{\input{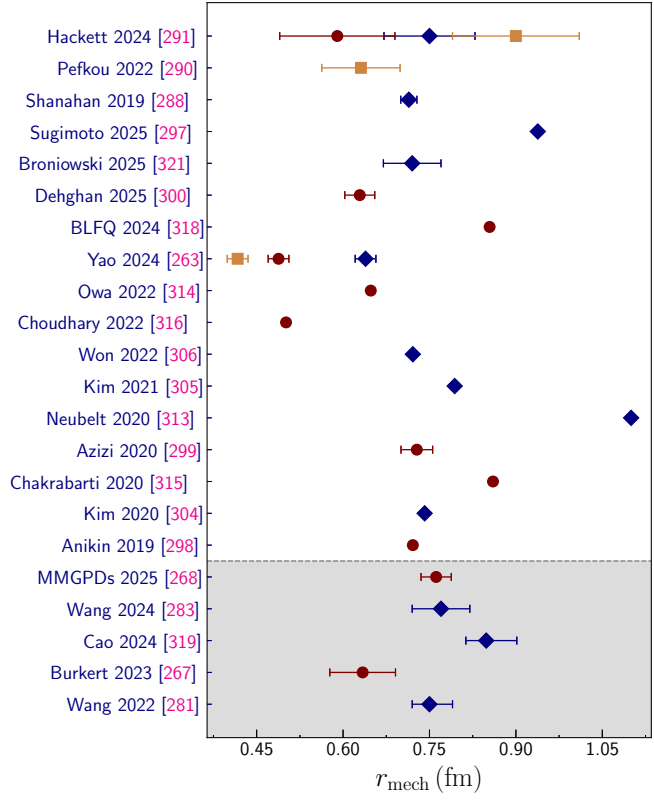}}
\caption{A comparison of the proton mechanical radius \( r_{\text{mech}} \) obtained from different theoretical techniques (top division) and various data analyses and phenomenological studies (bottom division). The results highlight the discrepancies and agreements among various methods. Circle, square, and diamond symbols represent the quark, gluon, and total contributions, respectively. See text for more information.}
\label{Fig:fig8}
\end{figure}

Figure~\ref{Fig:fig8} shows a comparison between the results obtained for the proton mechanical radius \( r_{\text{mech}} \). In this case, the results of theoretical techniques are presented in the top division of the figure, while the phenomenological results are shown in the bottom division. As before, the quark (circle), gluon (square), and total (diamond) contributions are presented separately for each study. As can be seen, there are fewer phenomenological studies determining \( r_{\text{mech}} \) compared to \( r_{\text{mass}} \). The phenomenological results span the interval \( 0.634 < r_{\text{mech}} < 0.849 \) fm, while the theoretical results cover \( 0.629 < r_{\text{mech}} < 1.1 \) fm. The combined values for the theoretical and phenomenological results are \( r_{\text{mech}} = 0.746 \pm 0.015 \) fm and \( r_{\text{mech}} = 0.753^{+0.021}_{-0.019} \) fm, respectively. Note that the combination process was performed as described earlier. Combining all phenomenological and theoretical results yields \( r_{\text{mech}} = 0.750^{+0.013}_{-0.012} \) fm, which is slightly smaller than the corresponding value obtained for \( r_{\text{mass}} \).
%

\section{Summary and conclusion}\label{sec:five} 
 
The study of nucleon radii provides critical insights into the internal structure of protons and neutrons, offering a window into the non-perturbative regime of QCD. In this work, we have comprehensively reviewed and analyzed the charge, magnetic, axial, mechanical, and mass radii of the nucleon, drawing from experimental, phenomenological, and theoretical approaches. Our findings highlight both the progress made in understanding the nucleon structure and the challenges that remain, particularly in reconciling discrepancies between different methods and interpretations.

The proton charge radius \( r_{pE} \) has been a focal point of intense study, especially in light of the proton radius puzzle, which revealed discrepancies between measurements from muonic hydrogen spectroscopy and traditional electron scattering or hydrogen spectroscopy. Our investigation shows that muonic hydrogen spectroscopy yields smaller values (\( r_{pE}= 0.84 \) fm) with remarkably high precision, while electron scattering and hydrogen spectroscopy results span a wider range. Overall, the experimental results cover the interval $ 0.831 < r_{pE} < 0.879 $ fm. The combined experimental value of \( r_{pE} = 0.857 \pm 0.005 \) fm aligns well with the phenomenological studies whose combination yields \( r_{pE} = 0.857 \pm 0.002 \) fm. Theoretical calculations, particularly from lattice QCD (\( r_{pE} = 0.815 \pm 0.016 \) fm), tend to predict smaller values (\( r_{pE} = 0.827 \pm 0.013 \) fm), highlighting the need for further refinement in both experimental and theoretical approaches. Combining all experimental, theoretical, and phenomenological results, we obtain \( r_{pE} = 0.847 \pm 0.005 \) fm.

For the proton magnetic radius \( r_{pM} \), the results show better agreement across different methods compared to the charge radius. The combined experimental and phenomenological values (\( r_{pM} = 0.846 \pm 0.017 \) fm and \( r_{pM} = 0.830 \pm 0.011 \) fm, respectively) are larger than the combined theoretical predictions (\( r_{pM} = 0.786 \pm 0.021 \) fm), suggesting that theoretical models may underestimate the spatial distribution of the nucleon's magnetic properties. Finally, combining all experimental, theoretical, and phenomenological results, we obtain \( r_{pM} = 0.821 \pm 0.01 \) fm.

The determination of the neutron's charge and magnetic radii is more challenging compared to the proton due to several intrinsic properties and experimental limitations. The corresponding combined values of the experimental, phenomenological, and theoretical results for the mean-square charge radius of the neutron, \( \left<r_{nE}^2\right> \), are \( \left<r_{nE}^2\right> = -0.115 \pm 0.002 \) fm\(^2\), \( \left<r_{nE}^2\right> = -0.110 \pm 0.005 \) fm\(^2\), and \( \left<r_{nE}^2\right> = -0.052 \pm 0.020 \) fm\(^2\), respectively. Combining all these results yields \( \left<r_{nE}^2\right> = -0.092 \pm 0.007 \) fm\(^2\).

The axial radius \( r_A \), which reflects the nucleon's weak charge distribution, has been studied through neutrino scattering, pion electroproduction, and theoretical frameworks such as lattice QCD and chiral perturbation theory. Our study reveals a wide range of values for \( \left<r_A^2\right> \), from 0.167 to 0.548 fm\(^2\), with combined experimental and phenomenological results (\( r_A = 0.645 \pm 0.101 \) fm and \( r_A = 0.628 \pm 0.093 \) fm, respectively) showing good consistency. Theoretical predictions, particularly from lattice QCD (\( r_A = 0.573 \pm 0.051 \) fm), tend to yield smaller values (\( r_A = 0.533^{+0.071}_{-0.072} \) fm), indicating a potential discrepancy that warrants further investigation.

The mechanical and mass radii provide unique insights into the spatial distribution of pressure, shear forces, and mass within the nucleon. For the proton mass radius \( r_{\text{mass}} \), theoretical calculations (\( r_{\text{mass}} = 0.727 \pm 0.025 \) fm) suggest smaller values compared to phenomenological studies (\( r_{\text{mass}} = 0.785 \pm 0.014 \) fm). Such a situation is also true for the combined theoretical and phenomenological values of the proton mechanical radius \( r_{\text{mech}} \) (\( r_{\text{mech}} = 0.746 \pm 0.015 \) fm and \( r_{\text{mech}} = 0.753^{+0.021}_{-0.019} \) fm, respectively). Combining all phenomenological and theoretical results yields \( r_{\text{mass}} = 0.756 \pm 0.014 \) fm and \( r_{\text{mech}} = 0.750^{+0.013}_{-0.012} \) fm. The slightly smaller value of \( r_{\text{mech}} \) compared to \( r_{\text{mass}} \) reflect differences in the spatial distribution of pressure and mass within the nucleon. Table I summarizes the combined experimental, lattice QCD, theoretical, and phenomenological results obtained in the present study for each radius separately. Note that, in this review, we classify lattice QCD calculations as theoretical results. While Table I provides separate combined values for lattice QCD predictions (for readers interested specifically in these results), these are included in the overall theoretical averages presented in the ``Theory" column. Additionally, the last column of the table includes the combination of all experimental, theoretical, and phenomenological results, providing a comprehensive overview of the nucleon's electromagnetic, weak, and mechanical properties. The combined values together with their uncertainties represent simple averages of independent results, obtained via standard procedure explained before, for illustrative purposes only. A rigorous statistical combination would require accounting for data overlaps and theoretical systematic differences, which is beyond the scope of this review. We also note that theoretical predictions differ substantially in their underlying assumptions and methodologies. While our simple averaging provides a compact overview, detailed comparisons should consider these fundamental differences on a case-by-case basis.
\begin{table}[th!]
\scriptsize
\setlength{\tabcolsep}{8pt} 
\renewcommand{\arraystretch}{1.4} 
\caption{Summary of combined experimental, lattice QCD, theoretical, and phenomenological results for different nucleon radii. Note that lattice QCD results are shown separately for convenience but are included in the overall Theory average. The last column provides the combination of all experimental, theoretical, and phenomenological results, offering a comprehensive overview of the nucleon's electromagnetic, weak, and mechanical properties. All values are given in femtometers (fm), except for the mean-square charge radius of the neutron \( \left< r_{nE}^2 \right> \), which is given in fm\(^2\).}\label{tab:par}
\begin{tabular}{lccccc}
\hline
\hline
 Quantities &  Experiment            &  Lattice QCD           &  Theory           &  Phenomenology            &   All\\
\hline 
\hline
$ r_{pE} $  & $ 0.857 \pm 0.005 $  & $ 0.815 \pm 0.016 $ & $ 0.827 \pm 0.013 $ & $ 0.857\pm 0.002 $ & $ 0.847 \pm 0.005 $  \\

$ r_{pM} $  & $ 0.846 \pm 0.017 $  & $ 0.759 \pm 0.040 $ & $ 0.786 \pm 0.021 $ & $ 0.830 \pm 0.011 $ & $ 0.821 \pm 0.010 $  \\

$ \left< r_{nE}^2 \right> $  & $ -0.115 \pm 0.002 $  & $ -0.075 \pm 0.013 $ & $ -0.076 \pm 0.020 $ & $ -0.110 \pm 0.005 $ & $ -0.100 \pm 0.007 $ \\

$ r_{nM} $  & $ 0.890 \pm 0.030 $  & $ 0.746 \pm 0.078 $ & $ 0.786 \pm 0.045 $ & $ 0.847 \pm 0.019 $ & $ 0.843\pm0.019 $  \\

$ r_A $  &  $ 0.645 \pm 0.101 $ & $ 0.573 \pm 0.051 $ & $ 0.533^{+0.071}_{-0.072} $ & $ 0.628 \pm 0.093 $  & $ 0.604 \pm 0.053 $  \\

$ r_{\text{mass}} $  & $ --- $  & $ 0.666 \pm 0.045 $ & $ 0.727 \pm 0.025 $ & $ 0.785 \pm 0.014 $ & $ 0.756 \pm 0.014 $ \\

$ r_{\text{mech}} $  & $ --- $  & $ 0.698 \pm 0.035 $ & $ 0.746 \pm 0.015 $ & $ 0.753^{+0.021}_{-0.019} $ & $ 0.750^{+0.013}_{-0.012} $ \\

\hline 		 	
\hline 	
\end{tabular}
\end{table}

According to the results obtained in the present study, several key findings and open questions can be raised: 

\begin{enumerate}
    \item The discrepancy between muonic hydrogen spectroscopy and electron scattering results remains unresolved, underscoring the need for improved experimental techniques and theoretical models.
    \item Lattice QCD and other theoretical approaches consistently predict smaller values for nucleon radii compared to experimental and phenomenological results, highlighting the importance of refining QCD-based calculations.
    \item The axial radius \( r_A \) and related weak interaction properties require further experimental data, particularly from neutrino scattering and parity-violating electron scattering, to reduce uncertainties and resolve discrepancies.
    \item The mechanical and mass radii provide complementary insights into the nucleon's internal structure, but the limited number of studies calls for more extensive investigations, particularly in the context of GPDs and lattice QCD.
    \item By synthesizing recent experimental results and theoretical advancements, we emphasize that each radius reflects a specific aspect of the nucleon's internal structure, such as its electric charge distribution, magnetic properties, weak interactions, or internal mechanical stress. Although the charge radius is often interpreted as the ``size" of the nucleon, this interpretation is an oversimplification that overlooks the multifaceted nature of nucleon structure.
     \item Table~\ref{tab:par} presents a summary of proton radius values and their quoted uncertainties from various approaches. We emphasize that determining a true world average with reliable uncertainty requires careful treatment of all systematic effects and correlations among different results, which is a considerable undertaking and not addressed here. Nevertheless, we have presented systematic comparisons of all available results in our figures. The apparent sub-percent uncertainties in some entries in Table~\ref{tab:par} reflect only the precision within individual studies, not the global knowledge of the proton radius.
\end{enumerate}

This work has provided a comprehensive overview of the different nucleon radii, including the charge and magnetic radii, the axial radius, as well as the mechanical and mass radii, emphasizing the progress made and the challenges that remain. We have addressed the common but misleading interpretation of the proton radius as a simple measure of its size, underscoring the nuanced and context-dependent nature of nucleon radii. We have tried to highlight the distinction between ``radii" (quantified via form factors) and ambiguous ``size" terminology often used in literatures. Considering the fact that the latter cannot even be properly defined in a field theory approach, a more focused inquiry from both experiment and theory is needed to address whether a universal nucleon radius exists. While significant advances have been achieved in both experimental and theoretical approaches, discrepancies such as the proton radius puzzle and the differences between theoretical and phenomenological results highlight the complexity of nucleon structure. Future experiments, such as those planned at Jefferson Lab, Mainz, and CERN, along with continued advancements in lattice QCD and other theoretical frameworks, will be crucial in achieving a unified understanding of the nucleon's internal structure. By addressing these challenges, we can deepen our understanding of QCD and its role in shaping the fundamental properties of matter.

%
\section*{ACKNOWLEDGEMENTS}
M.~Goharipour is thankful to the School of Particles and Accelerators and School of Physics, Institute for Research in Fundamental Sciences (IPM), for financial support provided for this research.  F. Irani and K. Azizi are tankful to Iran National Science Foundation (INSF) for financial support provided for this research under grant No. 4033039.

%

%

\bibliographystyle{apsrev4-1}
\bibliography{article} 

\end{document}